\documentclass[prl,
twocolumn,
preprintnumbers,showpacs,
superscriptaddress]{revtex4-1}
\usepackage{graphicx}
\usepackage{bm}
\usepackage{amsmath}
\usepackage{amssymb}
\usepackage[colorlinks,pdfstartview=FitH]{hyperref}
\usepackage{comment}
\hypersetup{linkcolor=blue,citecolor=blue,filecolor=black,urlcolor=blue}

\renewcommand\section[1]{\emph{#1}\ ---}

\newcommand\be{\begin{eqnarray}}
\newcommand\ee{\end{eqnarray}}

\begin{document}

\preprint{RBRC-1212}
\author{Dmitri Kharzeev}
\affiliation{Department of Physics and Astronomy, Stony Brook University,
Stony Brook, New York 11794-3800, USA}
\affiliation{Department of Physics, Brookhaven National Laboratory, Upton, NY 11973, USA}
\affiliation{RIKEN-BNL Research Center, Brookhaven National Laboratory,
Upton, New York 11973-5000, USA}
\author{Mikhail Stephanov}
\affiliation{Physics Department, University of Illinois at Chicago, Chicago, 
Illinois 60607, USA}
\author{Ho-Ung Yee}
\affiliation{Physics Department, University of Illinois at Chicago, Chicago, 
Illinois 60607, USA}
\affiliation{RIKEN-BNL Research Center, Brookhaven National Laboratory,
Upton, New York 11973-5000, USA}

\title{Anatomy of chiral magnetic effect in and out of equilibrium}

\begin{abstract}
We identify a new contribution to the chiral magnetic conductivity at
finite frequencies -- the magnetization current. This allows to
quantitatively reproduce the known field-theoretic time-dependent (AC) chiral magnetic response in terms of kinetic theory. We evaluate the corresponding AC chiral magnetic conductivity in two flavor QCD plasma at weak coupling. The magnetization current results from the spin response of chiral quasiparticles to magnetic field, and is thus proportional to the quasiparticle's $g$-factor.  In condensed matter systems, where the chiral quasi-particles are emergent and the $g$-factor can significantly differ from 2, this opens the possibility to tune the AC chiral magnetic response. 

\end{abstract}
\maketitle

\section{Introduction and summary}
The Chiral Magnetic Effect (CME) \cite{Fukushima:2008xe} (see \cite{Kharzeev:2013ffa,Kharzeev:2012ph,Miransky:2015ava,Kharzeev:2015znc,Landsteiner:2016led} for reviews and additional references) is the generation of electric current along an external magnetic field induced by the imbalance in the densities of right- and left-handed chiral fermions. It has been observed recently in Dirac \cite{Li:2014bha,Xiong:2015} and Weyl \cite{Huang:2015,Zhang:2015} semimetals. An evidence for CME in the quark-gluon plasma has been previously reported by high-energy heavy ion experiments at RHIC \cite{Abelev:2009ac,Abelev:2009ad,Adamczyk:2015eqo} and LHC \cite{Adam:2015vje}.

In the limit of a constant external magnetic field, the CME conductivity $\sigma_0^\chi$ is completely fixed by the chiral anomaly:
\be\label{cme}
{\bm j} = \frac{e^2}{2\pi^2} \mu_5 {\bm B} \equiv \sigma_0^\chi \ {\bm B},
\ee
where $\mu_5 = (\mu_R - \mu_L)/2$ is the chiral chemical potential describing the amount of imbalance between the densities of right- and left-handed fermions in the system.  Because it is topologically protected, the zero frequency CME conductivity $\sigma_0^\chi$ is universal and is not modified by interactions (even though the magnitude of magnetic field can be of course renormalized). As a result, the CME conductivity is reproduced in approaches that assume very different properties of the system, including perturbation theory (weak coupling) \cite{Fukushima:2008xe,Kharzeev:2009pj}, holographic correspondence (strong coupling) \cite{Yee:2009vw,Rebhan:2009vc,Gynther:2010ed}, hydrodynamics \cite{Son:2009tf,Kharzeev:2011ds}, and kinetic theory \cite{Son:2012wh,Stephanov:2012ki,Chen:2012ca,Son:2012zy,Basar:2013qia,Chen:2014cla,Monteiro:2015mea}.

However the frequency dependence of the CME conductivity $\sigma^\chi(\omega)$ in an oscillating magnetic field presents the following puzzle  \cite{Kharzeev:2009pj}: the real part of the conductivity drops from $\sigma_0^\chi$ to 
$\frac{1}{3} \sigma_0^\chi$ as soon as the frequency deviates from
$\omega = 0$. The computation in \cite{Kharzeev:2009pj} has been
performed using leading order perturbation theory, and thus assumed
the absence of interactions between the fermions. Taking account of
these interactions through the damping of fermion propagators
smoothens out the discontinuity in $\sigma^\chi(\omega)$
\cite{Satow:2014lva}, as conjectured in \cite{Kharzeev:2009pj}. In the
collisionless limit of \cite{Kharzeev:2009pj}, this perturbative
field-theoretic result has to be reproduced by the Chiral Kinetic
Theory (CKT) without the collision term -- and here there appears to be a problem, as we will now explain.

Indeed, kinetic theory operates with the single particle
distributions, and so the field-theoretic result of
\cite{Kharzeev:2009pj} may be interpreted in terms of the particle
energy shift $\Delta {\cal E}$ due to the interaction of chiral
particle's magnetic moment $\bm \mu={\bm p / 2 |\bm p|^2}$ with
magnetic field $\bm B$, $\Delta {\cal E}=-\bm \mu\cdot\bm B=-{\bm
  p\cdot\bm B\over 2 |\bm p|^2}$. In equilibrium, this energy shift
aligns $\bm \mu$ with $\bm B$ to lower the energy, which in turn
aligns the momentum $\bm p$ of chiral particles along $\bm B$. The
asymmetry between the left- and right-handed particles then results in
the electric current -- however, as will be shown below, this
alignment is responsible only for $1 \over 3$ of the result
(\ref{cme}). We will see that the rest $2 \over 3$ results from a more
kinematic effect including the Berry curvature $\bm b={\bm p\over
  2|\bm p|^3}$ of the monopole in momentum space  \cite{Volovik:2003fe}, that modifies both the phase space measure and the particle velocity. In fact, without Berry curvature, the kinematic modification of particle velocity due to the interaction with magnetic field yields a contribution to the current equal to $- {1 \over 3}$ of (\ref{cme}), and cancels exactly the contribution arising from the particle's energy shift - there is thus no CME in equilibrium without the Berry curvature. Since Berry curvature in momentum space is a quasi-classical description of chiral anomaly, this  explains why there is no CME in a system of chirally imbalanced fermions in the absence of anomaly \cite{Vilenkin:1980fu,Vilenkin:1980ft}.

When an external magnetic field oscillates with frequency $\omega$
that exceeds the inverse relaxation time $\tau_R$, $\omega >
\tau_R^{-1}$, the system is driven out of equilibrium, and the
contribution to CME conductivity from the energy shift in equilibrium
distributions should disappear. Since the leading order perturbative
result of \cite{Kharzeev:2009pj} corresponds to $\tau_R \to \infty$,
this means that as soon as $\omega$ deviates from zero, the CME
conductivity should only contain the contribution from the kinematic
effects including Berry curvature that does not require equilibrium
and hence survives at finite frequency. Based on this, we might expect that at small non-zero $\omega$ the CME conductivity should become $\frac{2}{3} \sigma_0^\chi$ -- in sharp contrast to $\frac{1}{3} \sigma_0^\chi$ computed in 
\cite{Kharzeev:2009pj}. 

What is the origin of this discrepancy between the field-theoretic and kinetic theory results?
It appears that in the kinetic theory description we miss a contribution to the CME conductivity that at finite frequency is negative and given by $- \frac{1}{3} \sigma_0^\chi$, and that vanishes in equilibrium, when $\omega =0$. 

In this paper we identify this missing contribution and show that it is
the magnetization current resulting from magnetization density $\bm m$ of chiral particles:
\be\label{mag_cur}
\bm J^M=\bm\nabla\times\bm m=-{\sigma_0\over 3}{\omega\over  \omega+i\tau_R^{-1}}\bm B  .
\ee
At frequencies $\omega \gg \tau_R^{-1}$ the magnetization current contributes an additional $-{1 \over 3}\sigma_0$ to the chiral magnetic conductivity $\sigma(\omega)$, whereas at zero frequency $\omega = 0$ this contribution vanishes. 

Since in chiral materials the Lorentz invariance is in general lost, the identification of individual contributions to the chiral magnetic conductivity at finite frequency may make it possible to manipulate them separately.

\vskip0.3cm

\section{CME in static equilibrium}
We start our discussion with a dissection of chiral magnetic
conductivity in static equilibrium. Throughout the paper, we shall
assume that the chirality relaxation time is long enough (in
particular, much longer than the typical relaxation time) that it
 makes sense to consider the chirally
imbalanced system as being close to equilibrium.

For brevity, we will consider a weakly interacting theory of single chiral (Weyl) fermion species of right-handed chirality; the generalization to a massless Dirac fermion (as well as to QCD in chiral limit) is straightforward. In this theory, the equilibrium CME current is given by $\bm J={\mu\over 4\pi^2}\bm B\equiv \sigma_0\bm B$ where $\mu$ is a chemical potential for the $U(1)$ chiral charge of the Weyl fermion.

In weakly coupled regime, a natural description of the system is in terms of a kinetic theory of quasi-particles described by the Boltzmann equation with collisions.
Collision terms that cause the system to relax to the thermal equilibrium set a characteristic relaxation time scale $\tau_R$ which, due to weak coupling, is much longer than the thermal quasi-particle scale, $\tau_R \gg T^{-1}$ where $T$ is the system's temperature. If the external background (such as the magnetic field $\bm B$) changes with a rate  slower than $\tau_R^{-1}$, we expect the system to stay in local thermal equilibrium; the hydrodynamic description of the system then applies. 
On the other hand, if an external field changes with a rate faster than $\tau_R^{-1}$, the collision terms become sub-leading and the Boltzmann equation reduces to the kinetic theory of non-interacting quasi-particles at leading order.
Therefore, $\tau_R^{-1}$ serves as a crossover scale between
hydrodynamic regime and the free streaming (collisionless) kinetic theory. In this section we will focus on CME in the hydrodynamic regime.

Quantization of a single right-handed Weyl field gives ``particles'' of helicity $h=+{1\over 2}$ and U(1) chiral charge $Q=+1$, as well as ``anti-particles'' carrying $h=-{1\over 2}$ and $Q=-1$. We will call them fermions and anti-fermions in the following, while the term ``quasi-particles'' will be used to refer to both. The helicity $h$ defines the spin of a
quasi-particle along its momentum direction as $\bm S=h\hat{\bm p}\equiv h{\bm p\over |\bm p|}$. Since the magnetic moment is given by $\bm \mu=Q{\bm S\over |\bm p|}$ (with the $g$ factor $g=2$), we see that the fermion and anti-fermion with a given momentum $\bm p$ carry the {\it same} magnetic moment $\bm \mu={\bm p\over 2 |\bm p|^2}$. In the presence of magnetic field $\bm B$, their energy is shifted by the same amount $\Delta {\cal E}=-\bm \mu\cdot\bm B=-{\bm p\cdot\bm B\over 2 |\bm p|^2}$ from the relativistic massless spectrum ${\cal E}_0=|\bm p|$. In equilibrium, this energy shift should give rise to a tendency of aligning $\bm \mu$ with $\bm B$ to lower the energy, which in turn aligns the momentum $\bm p$ of quasi-particles along $\bm B$: this has long been a qualitative explanation of CME in weakly coupled theory.

We will be more quantitative on this. The equilibrium thermal distributions of fermions ($f_+$) and anti-fermions ($f_-$) in the magnetic field $\bm B$ become
\be
f_{\pm}^{\rm eq}(\bm p)=f_\pm^{0}(|\bm p|-\Delta {\cal E})\approx f_\pm^0(|\bm p|)-\beta f_\pm^0(1-f_\pm^0)\Delta{\cal E},\label{eqshift}
\ee
where $f^0_\pm(x)=1/(e^{\beta (x\mp\mu)}+1)$ is the Fermi-Dirac distribution with a chemical potential $\mu$ and temperature $T \equiv \beta^{-1}$, that we have expanded to linear order in $\bm B$.
The U(1) charge current in kinetic theory (in leading order gradient expansion) is easy to understand intuitively: $\bm J=\int_{\bm p} \bm v_{\bm p}(f_+(\bm p)-f_-(\bm p))$, where $\int_{\bm p}\equiv \int {d^3\bm p\over (2\pi)^3}$ and $v_{\bm p}$ is the quasi-particle velocity which is $v_{\bm p}=\hat{\bm p}$ to leading order in $\bm B$ (we will come back to accounting the correction to velocity shortly). Then the net current resulting from the shift of distributions in (\ref{eqshift}) is
\be\label{eqCME}
\bm J^{\rm EQ}&=&{\beta\over 2}\int_{\bm p}{\bm p (\bm p\cdot\bm B)\over |\bm p|^3}\sum_{s=\pm}s f^0_s(|\bm p|)(1-f_s^0(|\bm p|))\nonumber\\
&=&{1\over 3}\times{\bm B\over 4\pi^2} \beta\int_0^\infty d|\bm p| |\bm p| \sum_{s=\pm}s f^0_s(|\bm p|)(1-f_s^0(|\bm p|))\nonumber\\
&=&{1\over 3}\times {\mu\over 4\pi^2}\bm B={1\over 3}\sigma_0\bm B\,,
\ee
where we have used an identity 
\be\label{ident1}
\beta\int_0^\infty d|\bm p| |\bm p| \sum_{s=\pm}s f^0_s(|\bm p|)(1-f_s^0(|\bm p|))=\mu
\ee
 that holds independently of temperature $\beta^{-1}=T$.
We see that the shift of thermal equilibrium distribution from the magnetic moment interaction  (\ref{eqshift}) explains only $1\over 3$ of the expected value of CME in equilibrium.

To identify the remaining $2\over 3$, we need a more careful description of kinetic theory of chiral fermions: the chiral kinetic theory.
A novel element in chiral kinetic theory is the inclusion of quantum effect originating from chiral spinors in first order of $\hbar\partial_x/p$, which appears
as a Berry curvature of monopole shape in {\it momentum} space: $\bm b={\bm p\over 2|\bm p|^3}$. This Berry curvature has been shown to be sufficient to describe the chiral anomaly in the framework of 
kinetic theory of quasi-particles \cite{Son:2012wh,Stephanov:2012ki,Chen:2012ca,Son:2012zy,Basar:2013qia,Chen:2014cla,Monteiro:2015mea}. We will now show that this Berry curvature plays a crucial role in explaining the rest  ${2\over 3}$ of the equilibrium CME conductivity.

The Berry curvature modifies the charge current $\bm J=\int_{\bm p} \sqrt{G} \bm v^B_{\bm p}(f_+(\bm p)-f_-(\bm p))$ in the following way: it affects the phase space measure $\sqrt{G}=(1+\bm b\cdot\bm B)$, and also changes the quasi-particle velocity $v_{\bm p}^B$:
\be
\sqrt{G}v_{\bm p}^B={\partial {\cal E}\over \partial\bm p}+\left({\partial {\cal E}\over \partial\bm p}\cdot\bm b\right)\bm B=\hat{\bm p}+{(\bm p\cdot\bm B)\bm p\over |\bm p|^4}+{\cal O}(\bm B^2)\,.\nonumber
\ee
To linear order in $\bm B$, the contribution to the net current arising from this modification of phase space and velocity (that is independent of the modification of the thermal equilibrium distribution considered above) can be obtained by using $f_\pm(\bm p)=f^0_\pm(|\bm p|)$ as 
\be\label{kinem}
\bm J^{\rm KM}&=&\int_{\bm p} {{\bm p}(\bm p\cdot\bm B)\over |\bm p|^4}(f_+^0(|\bm p|)-f^0_-(|\bm p|))\nonumber\\
&=&{2\over 3}\times {\bm B\over 4\pi^2}\int_0^\infty d|\bm p|(f_+^0(|\bm p|)-f^0_-(|\bm p|))\nonumber\\&=&
{2\over 3}\times {\mu\over 4\pi^2}\bm B={2\over 3}\sigma_0\bm B\,,
\ee
where
we use the identity $\int_0^\infty d|\bm p|(f_+^0(|\bm p|)-f^0_-(|\bm p|))=\mu$ that is related to the identity (\ref{ident1}) via integration by parts. Because this term in the current arises from the modification of quasiparticle velocity, we will call (\ref{kinem}) the {\it kinematic} contribution to CME; it is responsible for $2\over 3$ of the equilibrium CME conductivity.

If we had not included the Berry curvature, and simply used $v_{\bm p}^B={\partial {\cal E}\over\partial\bm p}$ with $\sqrt{G}=1$ (obtained by putting $\bm b=0$),
it is easy to check that we would get the kinematic contribution of $\bm J^{\rm KM}=-{1\over 3}\sigma_0\bm B$ that would cancel (\ref{eqCME}) resulting in the zero total CME in equilibrium. In other words, the change in the velocity $v_{\bm p}^B$ due to the energy shift from magnetic moment interaction with magnetic field precisely cancels the contribution from the shift of thermal equilibrium distribution due to the same energy shift in the equilibrium CME. We see that the net amount of the equilibrium CME, $\bm J=\sigma_0\bm B$, originates solely from the Berry curvature.

The dissection of $\bm J$ into $\bm J^{\rm EQ}$ and $\bm J^{\rm KM}$ described above is important both conceptually and practically when we consider out-of-equilibrium CME in free streaming regime discussed in the next section. In this regime, the thermal relaxation governed by the time scale $\tau_R$ cannot follow the fast change of the system induced by the time dependence of the background field. The contribution to CME from the shift of thermal equilibrium distribution $\bm J^{\rm EQ}$ is expected to be lost, while the kinematic contribution $\bm J^{\rm KM}$ that does not rely on relaxation dynamics should persist. We may thus naively expect that about ${2\over 3}$ of the equilibrium CME (carried by $\bm J^{\rm KM}$) will survive in out-of equilibrium conditions.
Interestingly, in the next section, we will see that the story of CME in out-of equilibrium is more subtle.

\vskip0.3cm

\section{Out-of equilibrium CME}
One way of driving the system out-of equilibrium is to apply a magnetic field oscillating in time with frequency $\omega$.
When the system reaches a stationary phase, the CME current oscillates with the same frequency and we can define a chiral magnetic conductivity $\sigma(\omega)$
by $\bm J(\omega)=\sigma(\omega)\bm B(\omega)$ in frequency space.
In the hydrodynamic regime $\omega\ll\tau_R^{-1}$, thermal relaxation dynamics is able to sustain instantaneous thermal equilibrium, and the
chiral magnetic conductivity should approach the equilibrium value $\sigma(\omega)\to\sigma_0$.
In the free-streaming regime $\omega\gg\tau_R^{-1}$ on the other hand, the system is unable to follow the rapidly oscillating magnetic field, and the distribution functions $f_\pm(\bm p)$ cannot deviate much from $f^0_\pm(|\bm p|)$. We would expect that only the kinematic contribution of CME, $\bm J^{\rm KM}$, manifests itself and that $\sigma(\omega)\to {2\over 3}\sigma_0$ in the free streaming regime.

A first hint of what should be truly happening in free streaming regime can be seen in the diagrammatic 1-loop computation of $\sigma(\omega)$ in Ref.\cite{Kharzeev:2009pj}, as well as in the Hard Thermal Loop (HTL) computation of Ref.\cite{Son:2012zy,Satow:2014lva,Manuel:2013zaa}.
The real part of $\sigma(\omega)$ at $\omega=0$ agrees with the equilibrium value $\sigma_0$ as expected, while it drops discontinuously to ${1\over 3}\sigma_0$
for an infinitesimally non-zero $\omega$. A careful inspection shows that this behavior of $\sigma(\omega)$ arises from the structure
\be
\sigma(\omega)=\sigma_0\left(1-{2\over 3}{\omega\over \omega+i\epsilon}\right)\,,\quad\epsilon=0^+\,,
\ee
which is a good approximation to the full $\sigma(\omega)$ in the kinetic HTL regime $\omega\ll T$. The presence of $i\epsilon$ is an artifact of non-interacting limit, which is replaced by the effective relaxation rate $i\tau_R^{-1}$ that smoothens the discontinuity at $\omega=0$ \cite{Satow:2014lva}. The important fact for us is that this result indicates $\sigma(\omega)\to {1\over 3}\sigma_0$ in the free streaming regime $\omega\gg \tau_R^{-1}$ after this replacement. As we mentioned in the Introduction, this is in puzzling contradiction to our expectation of $\sigma(\omega)\to {2\over 3}\sigma_0$ based on the dissection of equilibrium CME into $\bm J^{\rm EQ}$ and $\bm J^{\rm KM}$.
Since both $\bm J^{\rm EQ}$ and $\bm J^{\rm KM}$ follow from the structure of chiral kinetic theory, the only possible solution should be an existence of yet another contribution to CME that vanishes in hydrodynamic regime (i.e. in equilibrium) and approaches $-{1\over 3}\sigma_0\bm B$ in the free-streaming regime.
We will now show that such a contribution indeed exists.

The charge current we presented before, $\bm J=\int_{\bm p} \sqrt{G} \bm v^B_{\bm p}(f_+(\bm p)-f_-(\bm p))$ is the  leading classical term in the expansion in powers of $\hbar\partial_x/p$. Since the Berry curvature in momentum space is the first quantum correction in this parameter, we have to consider the first quantum corrections to the expression for $\bm J$. One such correction is a {\it magnetization current} arising from quantum spins of quasi-particles \cite{Chen:2014cla}: $\bm J^{\rm M}={\bm\nabla}\times\bm m$
where the magnetization density $\bm m$ is given by the sum of magnetic moments of quasi-particles $\bm \mu={\bm p\over 2|\bm p|^2}$ (recall that fermion and anti-fermion with a given momentum $\bm p$ carry the {\it same} magnetic moment), 
\be\label{magn}
\bm m=\int_{\bm p} {\bm p\over 2|\bm p|^2}(f_+(\bm p)+f_-(\bm p)).
\ee 
In fact, one can show that the current $\bm J^{\rm M}$ is responsible for the kinematic contribution of ${2\over 3}$ of the equilibrium chiral vortical effect (CVE) in the presence of non-zero fluid vorticity, and the rest ${1\over 3}$ of equilibrium CVE arises from the shift of thermal equilibrium distribution in the presence of vorticity \cite{Chen:2014cla}: this closely parallels the dissection of equilibrium CME we discuss before.
It turns out that this quantum current also gives rise to the additional contribution to out-of equilibrium CME, as we will now demonstrate.

The key element is the Bianchi identity, the well-known Faraday's Law
of induction, ${\partial\bm B\over\partial t}+\bm\nabla\times\bm E=0$
between a time-dependent magnetic field and a space-dependent electric
field, which in the frequency space becomes $-i\omega\bm
B+\bm\nabla\times\bm E=0$. Note that this is an identity in terms of
vector potential $(A_0,\bm A)$: the time-dependent magnetic field $i\omega\bm B$ is a space-dependent electric field $\bm\nabla\times\bm E$.
The electric field gives a local acceleration to quasi-particles $\dot{ \bm p}=Q\bm E=\pm\bm E$, and the free streaming Boltzmann equation in frequency space becomes
\be
{\partial f_\pm(\bm p)\over\partial t}+\dot{\bm p}\cdot{\partial f_\pm(\bm p)\over\partial\bm p}=-i\omega f_\pm(\bm p)\pm\bm E\cdot{\partial f_\pm(\bm p)\over\partial\bm p}=0\,.
\ee
The solution of this equation to linear order in $\bm E$ with a retarded boundary condition is $f_\pm(\bm p)=f_\pm^0(|\bm p|)+\delta f_\pm(\bm p)$ with 
$\delta f_\pm(\bm p)=\pm i {(\bm E\cdot\hat{\bm p})\over\omega+i\epsilon}\beta f^0_\pm(|\bm p|)(1-f^0_\pm(|\bm p|))$. The $i\epsilon$ 
should be replaced in the interacting theory by $i\tau_R^{-1}$ that appears in the collision term of the Boltzmann equation, that we neglect in free streaming regime.
The spin magnetization from this distribution is (see \ref{magn})
\be
\bm m&=&{i\beta\over\omega+i\tau_R^{-1}}\int_{\bm p} {\bm p\over 2|\bm p|^2}(\bm E\cdot\bm\hat{\bm p})\sum_{s=\pm} s f_s^0(|\bm p|)(1-f_s^0(|\bm p|))\nonumber
\\
&=&
{1\over 3}\times {i\bm E\over \omega+i\tau_R^{-1}}{\beta\over 4\pi^2}\int_0^\infty d|\bm p| |\bm p| \sum_{s=\pm}s f^0_s(|\bm p|)(1-f_s^0(|\bm p|))\nonumber
\\
&=&{1\over 3}\times {i\bm E\over \omega+i\tau_R^{-1}}\sigma_0\,.
\ee
This gives the magnetization current (\ref{mag_cur})
\be\label{mag_cur1}
\bm J^M=\bm\nabla\times\bm m={\sigma_0\over 3}{i(\bm\nabla\times \bm E)\over \omega+i\tau_R^{-1}}=-{\sigma_0\over 3}{\omega\over  \omega+i\tau_R^{-1}}\bm B 
\ee
 that contributes an additional $-{1 \over 3}\sigma_0$ to the chiral magnetic conductivity $\sigma(\omega)$ in 
 free-streaming regime $\omega\gg\tau_R^{-1}$: $\bm J^{\rm M}\to -{1\over 3}\sigma_0\bm B$. Together with $\bm J^{\rm KM}={2\over 3}\sigma_0\bm B$, this
explains the total CME current of $\bm J={1\over 3}\sigma_0\bm B$ in the free streaming regime.
Note that the physics of $\bm J^{\rm M}$ is independent of the Berry curvature.

\begin{table}[h]
\begin{tabular}{c|c|c}
 & $\omega\ll\tau_R^{-1}$& $\omega\gg\tau_R^{-1}$ \\
\hline
$\bm J^{\rm EQ}$ & ${1\over 3}$ & $0$ \\
$\bm J^{\rm KM}$ & ${2\over 3}$ & ${2\over 3}$\\
$\bm J^{\rm M}$ & $0$ & $-{1\over3}$\\\hline
$\bm J^{\rm total}$ & $1$ & ${1\over3}$ 
\end{tabular}
\caption{The dissection of CME in the hydrodynamic ($\omega\ll\tau_R^{-1}$) and the free-streaming ($\omega\gg\tau_R^{-1}$) regimes. The numbers are in units of the equilibrium CME conductivity.}
\end{table}

\section{Dependence on the g-factor}
The anatomy of CME in and out-of equilibrium presented above depends on the $g$-factor (taken to be $g=2$ for relativistic chiral fermions), and it
is easy to generalize our results to an arbitrary value of $g$-factor.
This discussion should be important in Dirac/Weyl semi-metals where the emergent pseudo-chiral fermions are expected to have a dynamically determined $g$-factor different from $g=2$. It also helps us to disentangle the physics of Berry curvature and the physics of spin magnetic moment ($g$-factor).
The magnetic moment of a quasi-particle of momentum $\bm p$ is $\bm \mu={g\over 2}Q{\bm S\over |\bm p|}$, and therefore the contribution from the shift of thermal distribution in equilibrium is modified by a factor of $g/2$, which becomes $\bm J^{\rm EQ}={g\over 6}\sigma_0\bm B$. It is easy to check that the effect of the Berry curvature together with the new energy shift gives
\be
\sqrt{G}v^B_{\bm p}&=&{\partial {\cal E}\over \partial\bm p}+\left({\partial {\cal E}\over \partial\bm p}\cdot\bm b\right)\bm B=\hat{\bm p}+{(2-g)\over 4}{\bm B\over |\bm p|^2}+{g\over 2}{(\bm p\cdot\bm B)\bm p\over |\bm p|^4},\nonumber
\ee
which results in the kinematic contribution to equilibrium CME as $\bm J^{\rm KM}=(1-{g\over 6})\sigma_0\bm B$. As expected, the total equilibrium CME, $\bm J = \bm J^{\rm EQ} + \bm J^{\rm KM} =\sigma_0\bm B$, is robust and does not depend on the value of  $g$-factor. 

In the free streaming regime $\omega\gg \tau_R^{-1}$, only the kinematic contribution of $\bm J^{\rm KM}=(1-{g\over 6})\sigma_0\bm B$ survives. On the other hand, the spin magnetization changes by a factor of ${g\over 2}$, therefore the magnetization current contribution will change by the same factor and become $\bm J^{\rm M}=-{g\over 6}\sigma_0\bm B$. The total CME in the free streaming regime is then $\bm J=\bm J^{\rm KM}+\bm J^{\rm M}=(1-{g\over 3})\sigma_0\bm B$. 

It is interesting to consider what the results would look like in the
absence Berry curvature (which can be seen by keeping only the terms with the $g$-factor).
The equilibrium CME is absent in this case, while the CME in free-streaming regime is $-{g\over 3}\sigma_0\bm B$.
Interestingly, the existence of CME in free streaming regime does not
necessarily require the Berry curvature, but its value would be
smaller by $\sigma_0\bm B$ from the one with the Berry
curvature. The importance of distinguishing the effects of the 
Berry curvature and of the magnetic moment has been also emphasized in the context of gyrotropic effect
\cite{Moore1,Moore2,Pesin}. We
emphasize again that the equilibrium CME in hydrodynamic regime, $\bm
J=\sigma_0 \bm B$, is a consequence of the Berry curvature,
independent of the physics of spin magnetic moment ($g$-factor).

\begin{table}[h]
\begin{tabular}{c|c|c}
 & $\omega\ll\tau_R^{-1}$& $\omega\gg\tau_R^{-1}$ \\
\hline
$\bm J^{\rm EQ}$ & ${g\over 6}$ & $0$ \\
$\bm J^{\rm KM}$ & $1-{g\over6}$ & $1-{g\over6}$\\
$\bm J^{\rm M}$ & $0$ & $-{g\over6}$\\\hline
$\bm J^{\rm total}$ & $1$ & $1-{g\over3}$ 
\end{tabular}
\caption{The dissection of CME with an arbitrary $g$-factor.
The terms with the $g$-factor originate from the physics of spin magnetic moment, and the others from the physics of Berry curvature.}
\end{table}

\vskip0.3cm

\section{Interpolating hydrodynamic regime and free streaming regime}
In interacting systems, such as QCD and Dirac/Weyl semi-metals in weakly coupled regime, the transition between the hydrodynamic regime and 
the free streaming regime should be well approximated by a function
\be
\sigma(\omega)=\sigma_0\left(1-{g\over 3}{\omega\over \omega+i\tau_R^{-1}}\right)\,,\label{inter}
\ee
with a single effective parameter $\tau_R$ with a dimension of time. One way to fix this parameter is to consider
a small frequency expansion, $\sigma(\omega)=\sigma_0-i\xi_5\omega+{\cal O}(\omega^2)$ with $\xi_5=-{g\over 3}\sigma_0\tau_R$,
where $\xi_5$ is one of the parity-odd transport coefficients in second order hydrodynamics, $\bm J=\sigma_0\bm B+\xi_5 {d\bm B\over dt}$ \cite{Kharzeev:2011ds}.
In two-flavor QCD, it has been computed in leading-log accuracy of coupling constant $\alpha_s\equiv g_s^2/(4\pi)$ by re-summing leading pinch singularities \cite{Jimenez-Alba:2015bia}, $\xi_5\approx -{0.5\over \alpha_s^2\log(1/\alpha_s) }{\sigma_0\over T}$, which gives (with $g=2$)
\be
\tau_R^{-1}\approx 1.3\, \alpha_s^2\log(1/\alpha_s) T\qquad({\rm 2-flavor\,\, QCD})\,.\label{tau}
\ee

We also mention that the imaginary part of $\sigma(\omega)$ is proportional to the parity-odd spectral density \cite{Jimenez-Alba:2015bia,Mamo:2015xkw} $\rho_{\rm odd}(k)=-2\,{\rm Im}\,\sigma(k)$ that
appears in the thermal fluctuation-dissipation relation of charge current $
\langle \bm J^i(k)\bm J^j(-k)\rangle\sim\left({1\over 2}+n_B(\omega)\right)i\epsilon^{ijl}\bm k^l\rho_{\rm odd}(k)$.
As can be seen from (\ref{inter}) the $\rho_{\rm odd}(\omega)$ is proportional to the $g$-factor -- this provides an intuitive explanation of why $\rho_{\rm odd}(k)$, though it arises from the same triangle diagram, is sensitive to the microscopic details of dynamics that are not constrained by the chiral anomaly.

\vskip0.3cm

\section{Towards a full quantum description of AC chiral magnetic conductivity in QCD}
To describe the AC CME conductivity $\sigma(\omega)$, we have to keep in mind  that the kinetic theory is valid only in the classical regime of quasi-particles, $\omega\ll T$, where $T$ characterizes the typical energy-momentum of thermal quasi-particles. For $\omega\gg T$ the probing scale is smaller than the Compton length of quasi-particles, and the response function is governed by quantum dynamics. This quantum response is dominated by a sum of one-particle responses of quasi-particles without collisions (since the scale of collisions is well separated from $T$, see (\ref{tau})), and it is reliably captured by a 1-loop diagrammatic computation \cite{Kharzeev:2009pj}. 

\begin{figure}
  \centering
  \includegraphics[height=4cm]{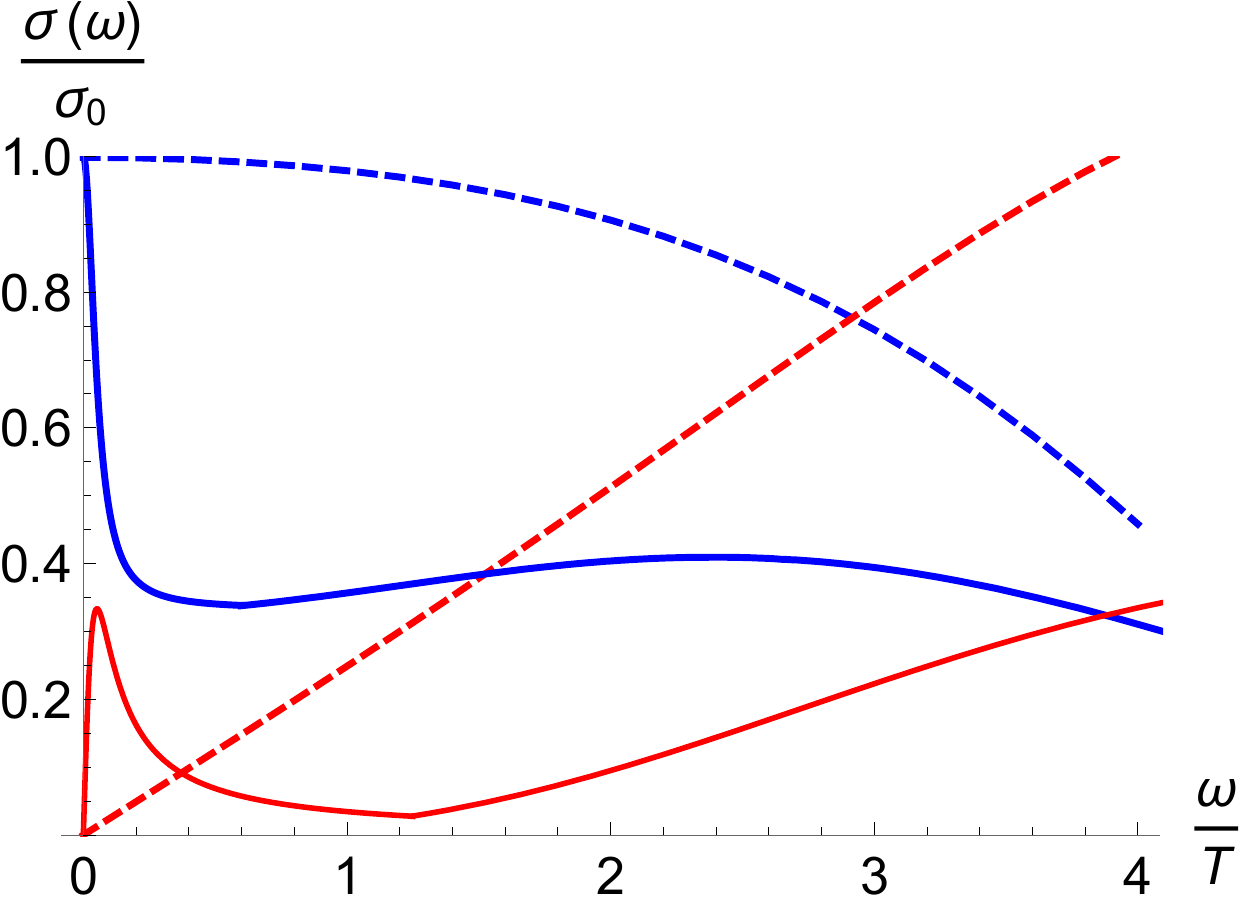}  \caption{The real (blue) and imaginary (red) parts of $\sigma(\omega)$ in two-flavor pQCD with $\alpha_s=0.2$ and $\mu_A/T=0.1$ (solid lines). The dotted curves are the results in the AdS/CFT correspondence \cite{Yee:2009vw}. }
  \label{fig1}
\end{figure}

This suggests a smooth interpolation of the previous kinetic theory result (\ref{inter}) in $\omega\ll T$ with a 1-loop diagrammatic result in $\omega\gg T$.
In FIG. \ref{fig1}, we plot the emerging global picture of $\sigma(\omega)$ in two-flavor perturbative QCD with $\alpha_s=0.2$ (neglecting the log factor) and $\mu_A/T=0.1$.
For comparison, we also show a result in the AdS/CFT correspondence for strongly coupled regime. One can see that at large frequency the pQCD and AdS/CFT results for the real part of conductivity approach each other, whereas at small frequency these results significantly differ signaling the difference in the relaxation mechanisms in these approaches. At zero frequency, the CME conductivity is completely fixed by the chiral anomaly and is universal. The imaginary part of AC conductivity is seen to exhibit different behavior both at small and large frequencies. 
\vskip0.3cm
\section{Discussion}
We have identified the new contribution to chiral magnetic conductivity -- the magnetization current (\ref{mag_cur1}) that allows to quantitatively reproduce the field-theoretic AC response in terms of kinetic theory. The magnetization current results from the spin response of chiral quasiparticles to magnetic field, and is thus proportional to the quasiparticle's $g$-factor, see Table II. 

The Lorentz-invariance of Dirac and Weyl actions fixes the value of $g$-factor to $g=2$, and the field-theoretic result automatically takes account of the corresponding magnetization current contribution. However, in Dirac/Weyl semimetals, the chiral quasiparticles are emergent, and the value of $g$-factor can significantly differ from $g=2$ -- for example, in ${\rm Bi_2Se_3}$ that possesses 3D Dirac quasiparticles with a finite mass of about $160-300$ meV the value of $g$-factor is $g=20-30$ \cite{Wolos} due to a strong spin-orbit interaction. Moreover, left- and right-handed chiral quasiparticles may have different properties in so-called asymmetric Weyl semimetals, leading to essential modification of CME \cite{Kharzeev:2016mvi}. 

Therefore, the identification of individual contributions to CME conductivity presented above opens the possibility to tune the chiral magnetic response at finite frequencies by choosing the appropriate chiral materials.

\vskip0.3cm
\section{Acknowledgment} We thank Qiang Li, Larry McLerran, Dam Son and Naoki Yamamoto for useful comments.
This work is partially supported by the U.S. Department of Energy, Office of Science, Office of Nuclear Physics, within the framework of the Beam Energy Scan Theory (BEST) Topical Collaboration, and grant No.\ DE-FG0201ER41195.

\bibliographystyle{my-refs}

\end{document}